\documentclass[proceedings]{JHEP3}
\usepackage[T1]{fontenc}

\makeatletter

\providecommand{\LyX}{L\kern-.1667em\lower.25em\hbox{Y}\kern-.125emX\@}

\PrHEP{PrHEP unesp2002}                 
\conference{Workshop on Integrable Theories, Solitons and Duality}

\title{BPS $Z_k$ strings, string tensions and confinement in 
non-Abelian theories}
 
\author{\speaker{Marco A. C. Kneipp} \\                      
Universidade do Estado do Rio de Janeiro(UERJ),\\
Depto. de Física Teórica,\\
Rua São Francisco Xavier, 524\\
20550-013, Rio de Janeiro, Brazil.\\
\vskip 0.1cm
Centro Brasileiro de Pesquisas F\' \i sicas (CBPF), \\ 
Coordenação de Teoria de Campos e Partículas (CCP), \\ 
Rua Dr. Xavier Sigaud, 150 \\
22290-180, Rio de Janeiro, Brazil.   \\                                  
         E-mail: \email{kneipp@cbpf.br}}                       

\abstract{In this talk we  review some generalizations of 
't Hooft and Mandelstam ideas on confinement for  theories with non-Abelian 
unbroken gauge groups. In order to do that, we consider N=2 super Yang-Mills 
with one flavor and a mass breaking term. One of the spontaneous 
symmetry breaking 
is accomplished   by a scalar  that can be in particular in the representation of 
the diquark condensate. We  analyze   the phases of the theory. 
In the superconducting phase, we   show the existence of BPS 
$Z_k$-strings and calculate exactly their string tension in a straightforward 
way. We also find  that magnetic fluxes of the monopole and $Z_k$-strings  
are proportional to one another allowing for monopole confinement in a 
phase transition. We  
further show that some of the resulting confining theories can be obtained 
by adding a deformation term to $N=2$ or $N=4$ superconformal theories.
 }
\makeatother

\begin{document}

\section{Introduction}

One of the oldest open problems in particle physics is the quark confinement.
It is believed that it could be explained as being a phenomena dual to a non-Abelian
generalization of Meissner effect, as was proposed by 't Hooft and Mandelstam
many years ago \cite{tHooftMandelstam}. Some progress has been made by Seiberg
and Witten \cite{SeibergWitten1}, who considered an \( N=2 \) \( SU(2) \)
supersymmetric theory and associated to the point in the moduli space where
the monopole becomes massless they obtained an effective \( N=2 \) super QED
with an \( N=2 \) mass breaking term. In this effective theory, the \( U(1) \)
is broken to a discrete group and as it happens the theory develops string solutions
and confinement of (Abelian) electric charges occurs. Since then, many vary
interesting works appeared\cite{varios}. However, usually it is considered
that the gauge group is completely broken to \( U(1)^{\textrm{rank}(G)} \)
and then to its discrete center by Higgs mechanism. Therefore these theories
don't have \( SU(3)\times U(1)_{\textrm{em}} \) as subgroup of the unbroken
gauge group and the monopoles belong to \( U(1) \) representations and not
to representations of non-Abelian groups.

In this talk, we shall review \cite{KB} and \cite{Kneipp2002} where we generalize
some of the 't Hooft and Mandelstam ideas to non-Abelian theories but avoiding
the mentioned problems. In order to do that, we consider \( N=2 \) super Yang-Mills
with a breaking mass term and with arbitrary simple gauge group which is broken
to non-Abelian residual gauge group. One of the spontaneous symmetry breaking
is produced by a complex scalar \( \phi  \) that could be for example in the
symmetric part of the tensor product of \( k \) fundamental representations.
In particular if \( k=2 \), this scalar is in the representation of a diquark
condensate and therefore it can be thought as being itself the condensate. We
therefore could consider this theory as being an effective theory (like the
\( N=2 \) \( SQED \) considered by Seiberg and Witten). The non-vanishing
expectation value of \( \phi  \) gives rise to a monopole confinement, like
in the Abelian-Higgs theory. We shall show that, by varying a mass parameter
\( m \), we can pass from an unbroken phase to a phase with free monopoles
and then to a superconducting phase with \( Z_{k} \)-strings and confined monopoles.
In the free-monopole phase, there exist (solitonic) monopole solutions which
are expected to fill irreducible representations of the dual unbroken gauge
group\cite{HolloDoFraMACK}. In this phase we recover \( N=2 \) supersymmetry
and show that some of these theories are conformal invariant. In the superconducting
phase we shall prove the existence of BPS \( Z_{k} \)-string solutions and
calculate exactly their string tensions. We also show that the fluxes of the
magnetic monopoles and strings are proportional to one another and therefore
the monopoles can get confined. From the values of the magnetic fluxes we calculate
the threshold length for the string breaking, producing a new monopole-antimonopole
pair. In our theory the bare mass \( \mu  \) of \( \phi  \) is not required
to satisfy \( \mu ^{2}<0 \) in order to have spontaneous symmetry breaking.
Therefore in the dual formulation, where one could interpret \( \phi  \) as
the monopole condensate, when \( k=2 \), we don't need to have a monopole mass
satisfying the problematic condition \( M_{\textrm{mon}}^{2}<0 \) mentioned
by 't Hooft\cite{tHooftreview}.

\section{Confinement in Abelian-Higgs theory}

Due to the broad audience in this conference, let us review the BPS string solutions
in Abelian-Higgs theory and the basic ideas of 't Hooft and Mandelstam on confinement\footnote{%
For a review in those subjects see \cite{GM(86)} and \cite{tHooftreview}.
}. 

As is well known, a superconductor is described by the BCS theory. In this theory
it happens a condensation of electron pairs. This condensate is associated to
a complex scalar field whose dynamics is governed by the Abelian-Higgs Lagrangian

\begin{equation}
\label{1.1}
L=-\frac{1}{4}F_{\mu \nu }F^{\mu \nu }+\frac{1}{2}D_{\mu }\phi ^{*}D^{\mu }\phi -V(\phi )\, ,
\end{equation}
 where \( D_{\mu }\phi =\partial _{\mu }\phi +iq_{\phi }A_{\mu }\phi  \) and
\[
V(\phi )=\frac{\lambda }{8}\left( |\phi |^{2}-a^{2}\right) \, .\]
The constant \( q_{\phi } \) is the electric charge of \( \phi  \). In particular
if \( \phi ^{\dagger } \) is a condensate of electron pairs, then \( q_{\phi }=2e \).
In this case, when \( a^{2}>0 \), the \( U(1) \) gauge group is broken to
\( Z_{2} \). In that phase, Abelian-Higgs is the effective theory which describes
normal superconductors. Since in this phase \( \Pi _{1}(U(1)/Z_{2}) \) is nontrivial
we can have string solutions. In order to obtain these solutions we shall look
for a static configuration, with cylindrical symmetry around the \( z \)-axis
and the only non-vanishing component of the field strength is \( B_{3}\equiv -F_{12} \).
In order that the string have a finite string tension \( T \) (i.e. energy
per unit length), when the radial coordinate \( \rho \rightarrow \infty  \),
the string solution must satisfy the vacuum equations
\begin{eqnarray}
D_{\mu }\phi  & = & 0\, ,\nonumber \\
V(\phi ) & = & 0\, ,\label{2.3} \\
F_{\mu \nu } & = & 0\, .\nonumber 
\end{eqnarray}
 Then, one can obtain the string tension lower bound \cite{Bog(76)} (for a
review see \cite{GM(86)})
\begin{equation}
\label{2.3a}
T\geq \frac{1}{2}q_{\phi }a^{2}\left| \Phi _{\textrm{st}}\right| \, ,
\end{equation}
 where
\[
\Phi _{\textrm{st}}\equiv \int d^{2}x\, B_{3}=-\oint dl_{I\, }A_{I}\, ,\, \, \, I=1,2\, ,\]
 is the string magnetic flux. We shall adopt the convention that capital Latin
indices always denote the coordinates \( I=1,2 \). From the boundary conditions
(\ref{2.3}), it follows that at \( \rho \rightarrow \infty  \)
\begin{eqnarray}
|\phi |=a & \, \rightarrow \,  & \phi (\varphi )=ae^{i\beta (\varphi )}\, ,\label{1.3} \\
D_{I}\phi =0 & \, \rightarrow \,  & A_{I}(\varphi )=\frac{i}{q_{\phi }}\phi ^{-1}\partial _{I}\phi =\frac{\epsilon _{IJ}x^{J}}{q_{\phi }\rho ^{2}}\partial _{\varphi }\beta \, ,\nonumber \label{1.3} 
\end{eqnarray}
where \( \beta (\varphi ) \) can be a multi-valued function. But since \( \phi (\varphi ) \)
is single valued 
\[
\beta (\varphi +2\pi )-\beta (\varphi )=2\pi n\, \, ,\, \, \, \mbox {where}\, \, \, n\in Z\, .\]
Then, 
\begin{equation}
\label{fluxo}
\Phi _{\textrm{st}}=\frac{2\pi }{q_{\phi }}n\, ,
\end{equation}
and it results that
\begin{equation}
\label{1.4}
T\geq a^{2}\pi \left| n\right| \, .
\end{equation}
The bound is saturated when\cite{Bog(76)}
\begin{eqnarray}
D_{0}\phi \, =\, D_{3}\phi  & = & 0\, ,\nonumber \label{0.2} \\
D_{\pm }\phi  & = & 0\, ,\label{1.5} \\
B_{3}\pm \frac{q_{\phi }}{2}\left( \phi ^{*}\phi -a^{2}\right)  & = & 0\, ,\nonumber \\
V(\phi ) & = & \frac{q_{\phi }^{2}}{8}\left( |\phi |^{2}-a^{2}\right) ^{2}\nonumber 
\end{eqnarray}
with ``\( + \)'' if \( n>0 \) and ``\( - \)'' if \( n<0 \) and where
\( D_{\pm }\equiv D_{1}\pm iD_{2} \). The last relation implies the constraint
on the couplings of the theory \( \lambda =q_{\phi }^{2} \). This constraint
appear in \( N=2 \) super-Maxwell. In order for this Lagrangian to be \( N=2 \)
super-Maxwell we just need to introduce some extra fields (note that since \( \phi  \)
has a non-vanishing electric charge it should belong to the hypermultiplet and
not to the \( N=2 \) vector supermultiplet). One can show that the solutions
of these equations satisfy automatically the equations of motion. 

The string ansatz is constructed by multiplying the asymptotic configuration
(\ref{1.3}) by arbitrary functions of \( \rho  \)\cite{ANO},
\begin{eqnarray*}
\phi (\varphi ,\rho ) & = & f(\rho )ae^{in\varphi }\, ,\\
A_{I}(\varphi ,\rho ) & = & g(\rho )\frac{n}{q_{\phi }\rho ^{2}}\epsilon _{IJ}x^{J}\, ,
\end{eqnarray*}
 and, in order to recover the asymptotic configuration at \( \rho \rightarrow \infty  \),
we consider the boundary condition 
\[
g(\infty )=1=f(\infty )\, .\]
On the other hand, in other to kill the singularities at the origin, we consider
the boundary condition 
\[
g(0)=0\textrm{ and }f(0)=0.\]
 Putting this ansatz in the BPS conditions, it results the first order differential
equations
\begin{eqnarray}
g'(\rho ) & = & \mp \frac{q_{\phi }^{2}a^{2}\rho }{2n}\left( |f(\rho )|^{2}-1\right) \, ,\label{0.4} \\
f'(\rho ) & = & \pm \frac{n}{\rho }\left( 1-g(\rho )\right) f(\rho )\, .\nonumber 
\end{eqnarray}
Although they don't have an analytic solution, Taubes has proven \cite{taubes}
the existence of a solution to these differential equations with the above boundary
conditions. Moreover, he showed that all solutions to the full static equations
are solutions to BPS equations, if \( \lambda =q^{2}_{\phi } \). Since that
BPS string solution has the lowest value of the string tension in a given topological
sector, it is automatically stable.

't Hooft and Mandelstam \cite{tHooftMandelstam} had the idea that if one puts
a (Dirac) monopole and antimonopole in a superconductor, their magnetic lines
could not spread over space but must rather form a string which gives rise to
a confining potential between the monopoles. This idea only makes sense since
the (Dirac) monopole magnetic flux is 
\[
\Phi _{\textrm{mon}}=g=2\pi /e\, ,\]
which is consistent with the string's magnetic flux quantization condition (\ref{fluxo}),
allowing one to attach to the monopole two strings with \( n=1 \), when \( q_{\phi }=2e \).
Then, using the electromagnetic duality of Maxwell theory one could map this
monopole confining system to an electric charge confining system.

\section{The BPS conditions for strings in non-Abelian theories}

Let us now generalize some of these ideas to a non-Abelian theory\cite{KB}\cite{Kneipp2002}.
For simplicity, let us consider an arbitrary gauge group \( G \) which is simple,
connected and simply-connected. We shall consider a Yang-Mills theory with a
complex scalar \( S \) in the adjoint representation and another complex scalar
\( \phi  \). We consider a scalar \( S \) in the adjoint representation because
in a spontaneous symmetry breaking it produces an exact symmetry group \( G_{S} \)
with a \( U(1) \) factor, which allows the existence of monopole solutions.
Additionally, another motivation for having a scalar in the adjoint representation
is because with it, we can form an \( N=2 \) vector supermultiplet and, like
in the Abelian-Higgs theory, the BPS string solutions appear naturally in a
theory with \( N=2 \) supersymmetry. Moreover, in a theory with particle content
of \( N=2 \) supersymmetry, the monopole spin is consistent with the quark-monopole
duality\cite{Osborn} which is another important ingredient in 't Hooft and
Mandelstam's ideas. However, with \( S \) in the adjoint, we can not in general
produce a spontaneous symmetry breaking which has a non-trivial first homotopy
group of the vacuum manifold, which is a necessary condition for the existence
of a string. One way to produce a spontaneous symmetry breaking satisfying this
condition is to introduce a complex scalar \( \phi  \) in a representation
which contains the weight state \( \left| k\lambda _{\phi }\right\rangle  \)
\cite{OT}, where \( k \) is an integer greater or equal to two, and \( \lambda _{\phi } \)
a fundamental weight. We can have at least three possibilities: one is to consider
\( \phi  \) in the representation with \( k\lambda _{\phi } \) as highest
weight, which we shall denote \( R_{k\lambda _{\phi }} \). We can also consider
\( \phi  \) to be in the direct product of \( k \) fundamental representations
with fundamental weight \( \lambda _{\phi } \), which we shall denote \( R_{k\lambda _{\phi }}^{\otimes } \).
Finally a third possibility would be to consider \( \phi  \) in the symmetric
part of \( R_{k\lambda _{\phi }}^{\otimes } \), called \( R_{k\lambda _{\phi }}^{\textrm{sym}} \),
which always contains \( R_{k\lambda _{\phi }} \). This last possibility has
an extra physical motivation that if \( k=2 \), it corresponds to the representation
of a condensate of two fermions (which we naively will call quarks) in the fundamental
representation with fundamental weight \( \lambda _{\phi } \), and we can interpret
\( \phi  \) as being this diquark condensate, similarly to the Abelian theory.
In this case, when \( \phi  \) takes a non-trivial expectation value, it also
gives rise to a mass term for these quarks. We shall see that \( \phi  \) will
be responsible for the monopole confinement. Therefore, in this case we shall
have that a nonzero vacuum expectation value of the diquark condensate gives
rise to monopole confinement in a non-Abelian theory. If we are considering
the dual theory, \( \phi  \) could be interpreted as a monopole condensate
and it would give rise to a quark confinement.

In order to have \( N=2 \) supersymmetry, we should need another complex scalar
to be in the same hypermultiplet as \( \phi  \). For simplicity's sake, however,
we shall ignore it setting it to zero.

Let us then consider the Lagrangian
\begin{equation}
\label{1.2}
L=-\frac{1}{4}G^{\mu \nu }_{a}G_{a\mu \nu }+\frac{1}{2}\left( D_{\mu }S\right) ^{*}_{a}\left( D^{\mu }S\right) _{a}+\frac{1}{2}\left( D_{\mu }\phi ^{\dagger }\right) \left( D^{\mu }\phi \right) -V(S,\phi )
\end{equation}
with potential given by
\[
V(S,\phi )=\frac{1}{2}\left( Y_{a}Y_{a}+F^{\dagger }F\right) \]
where 
\begin{eqnarray*}
Y_{a} & = & \frac{e}{2}\left( \phi ^{\dagger }T_{a}\phi +S_{b}^{*}if_{abc}S_{c}-m\left( \frac{S_{a}+S_{a}^{*}}{2}\right) \right) \, ,\\
F & = & e\left( S^{\dagger }-\frac{\mu }{e}\right) \phi \, .
\end{eqnarray*}
This potential is the bosonic part of \( N=2 \) super Yang-Mills with one flavor
(when one of the aforementioned scalars of the hypermultiplet is put equal to
zero). The parameter \( \mu  \) gives a bare mass to \( \phi  \) and \( m \)
gives a bare mass to the real part of \( S \) which softly breaks \( N=2 \)
SUSY. The parameter \( m \) also is responsible for spontaneous gauge symmetry
breaking and, as for the mass parameter \( a \) in the Abelian case, one can
consider it as a function of temperature. In \cite{KB}, we started with a generic
potential and have shown that in order to obtain the string BPS conditions,
the potential is constrained to have this form with \( N=2 \) SUSY like in
the Abelian case.

In order that the string tension \( T \) be finite, the string configuration
at \( \rho \rightarrow \infty  \) must satisfy the vacuum equations
\begin{eqnarray}
D_{\mu }S=D_{\mu }\phi  & = & 0\, ,\nonumber \\
V(S,\phi ) & = & 0\, ,\label{7.3} \\
G_{\mu \nu } & = & 0\, .\nonumber 
\end{eqnarray}
 Let \( S=M+iN \), where \( M \) and \( N \) are real scalar fields and \( B^{a}\equiv B^{a}_{3}=-G^{a}_{12}/2 \).
Then one can show \cite{KB} that the string tension in this theory satisfy
the inequality
\begin{equation}
\label{7.5}
T\geq \frac{me}{2}\left| \int d^{2}x\left\{ M_{a}B_{a}\right\} \right| 
\end{equation}
and the bound is saturated if and only if
\begin{eqnarray}
D_{0}\phi  & = & D_{3}\phi =D_{0}S=D_{3}S=0\nonumber \label{7.6aa} \\
D_{\pm }\phi  & = & 0\, ,\nonumber \label{7.6a} \\
D_{\mp }S & = & 0\, ,\label{7.6c} \\
B_{a}\pm Y_{a} & = & 0\, ,\nonumber \label{7.6d} \\
V(S,\phi )-\frac{1}{2}Y_{a}^{2} & = & 0\, ,\nonumber \label{7.6e} 
\end{eqnarray}
which are BPS conditions for the string. Since it is a static configuration
with axial symmetry, the first conditions imply that \( W_{0}=0=W_{3} \). The
last BPS condition implies that \( F=0 \). One can check that 1/4 of the \( N=2 \)
supersymmetry transformations vanish for field configurations satisfying the
string BPS conditions in the limit \( m\rightarrow 0 \).

Differently from the Abelian case, these equations are only consistent with
the equations of motion when \( m \) vanishes \cite{KB}. However this condition
must be understood in the limiting case \( m\rightarrow 0 \), as we shall discuss
later on. Therefore, it is only in this limit that we can have BPS strings satisfying
(\ref{7.6a}). The explanation for this fact is the following: remember that
a static solution of the equations of motion correspond to an extreme of the
Hamiltonian, for solutions where \( W_{0}=0 \). On the other hand, a solution
of the BPS conditions saturate the bound of the string tension in a given topological
sector. But that does not necessarily guarantee that it is an extreme. 

Now we shall analyze if the theory have a vacuum which produces a spontaneous
symmetry consistent with the existence of string solution.

\section{Phases of the theory}

The vacuum equation \( V(S,\phi )=0 \) is equivalent to 
\begin{equation}
\label{2.4}
Y_{a}=0=F\, .
\end{equation}
In order to the topological string solutions to exist, we look for vacuum solutions
of the form 
\begin{eqnarray}
\phi ^{\textrm{vac}} & = & a|k\lambda _{\phi }>\, ,\nonumber \label{2.5a} \\
S^{\textrm{vac}} & = & b\lambda _{\phi }\cdot H\, ,\label{2.5b} \\
W^{\textrm{vac}}_{\mu } & = & 0\, ,\nonumber 
\end{eqnarray}
where \( a \) and \( b \) are complex constants, \( k \) is a integer greater
or equal to two and \( \lambda _{\phi } \) is an arbitrary fundamental weight.
If \( a\neq 0 \), this configuration breaks \( G\rightarrow G_{\phi } \) in
such a way that\cite{OT} \( \Pi _{1}(G/G_{\phi })=Z_{k} \), which is a necessary
condition for the existence of \( Z_{k}- \)strings. Let us consider that \( \mu >0 \).
Following \cite{KB}, from the vacuum conditions \( Y_{a}=0=F \), one can conclude
that 
\begin{eqnarray*}
|a|^{2} & = & \frac{mb}{k}\, ,\\
\left( kb\lambda _{\phi }^{2}-\frac{\mu }{e}\right) a & = & 0\, .
\end{eqnarray*}
 There are three possibilities:

\begin{description}
\item [(i)]If \( m<0\, \, \Rightarrow \, \, a=0=b \) and the gauge group \( G \)
remains unbroken.
\item [(ii)]If \( m=0\, \, \Rightarrow \, \, a=0 \) and \( b \) can be any constant.
In this case, \( S^{\textrm{vac}} \) breaks\cite{OT}
\begin{equation}
\label{2.6}
G\rightarrow G_{S}\equiv \left( K\times U(1)\right) /Z_{l}\, ,
\end{equation}
 where \( K \) is the subgroup of \( G \) associated to the algebra whose
Dynkin diagram is given by removing the dot corresponding to \( \lambda _{\phi } \)
from that of \( G. \) The \( U(1) \) factor is generated by \( \lambda _{\phi }\cdot H \)
and \( Z_{l} \) is a discrete subgroup of \( U(1) \) and \( K \). The \( N=2 \)
supersymmetry is restored in this case.
\item [(iii)]If \( m>0\, \Rightarrow  \)
\begin{eqnarray}
|a|^{2} & = & \frac{m\mu }{k^{2}e\lambda _{\phi }^{2}}\, ,\label{2.7a} \\
b & = & \frac{\mu }{ke\lambda _{\phi }^{2}}\, ,\label{2.7b} 
\end{eqnarray}
and \( G \) is further broken to\cite{OT}
\begin{equation}
\label{2.8}
G\rightarrow G_{\phi }\equiv \left( K\times Z_{kl}\right) /Z_{l}\supset G_{S}\, .
\end{equation}
In particular for \( k=2 \) we have for example, 
\begin{eqnarray*}
\textrm{Spin}(10) & \rightarrow  & \left( SU(5)\times Z_{10}\right) /Z_{5}\, ,\\
SU(3) & \rightarrow  & \left( SU(2)\times Z_{4}\right) /Z_{2}\, .
\end{eqnarray*}

\end{description}
Therefore by continuously changing the value of the parameter \( m \) we can
produce a symmetry breaking pattern \( G\, \rightarrow \, G_{S}\, \rightarrow \, G_{\phi } \).
It is interesting to note that, unlike the Abelian-Higgs theory, in our theory
the bare mass \( \mu  \) of \( \phi  \) is not required to satisfy \( \mu ^{2}<0 \)
in order to have spontaneous symmetry breaking. Therefore in the dual formulation,
where one could interpret \( \phi  \) as the monopole condensate when \( \phi \in R_{2\lambda _{\phi }}^{\textrm{sym}} \),
we don't need to have a monopole mass satisfying the problematic condition \( M_{\textrm{mon}}^{2}<0 \)
mentioned by 't Hooft\cite{tHooftreview}. 

Let us analyze in more detail the last two phases\cite{Kneipp2002}.

\section{The \protect\( m=0\protect \) or free-monopole phase}

Since \( b \) is an arbitrary non-vanishing constant, we shall consider it
to be given by (\ref{2.7b}), in order to have the same value as the case when
\( m<0 \). The non-vanishing expectation value of \( S^{\textrm{vac}} \) defines
the \( U(1) \) direction in \( G_{S} \), (\ref{2.6}), and one can define
corresponding \( U(1) \) charge as \cite{GOrev}
\begin{equation}
\label{3.1}
Q\equiv e\frac{S^{\textrm{vac}}}{|S^{\textrm{vac}}|}=e\frac{\lambda _{\phi }\cdot H}{|\lambda _{\phi }|}\, .
\end{equation}
 Since 
\[
Q\phi ^{\textrm{vac}}=ek|\lambda _{\phi }|\, \phi ^{\textrm{vac}}\, ,\]
the electric charge of \( \phi ^{\textrm{vac}} \) is 
\begin{equation}
\label{4.1}
q_{\phi }=ek|\lambda _{\phi }|\, .
\end{equation}

Since in this phase \( \Pi _{2}(G/G_{S})=Z \), it can exist \( Z \)-magnetic
monopoles. Indeed, for each root \( \alpha  \) such that \( 2\alpha ^{\textrm{v}}\cdot \lambda _{\phi }\neq 0 \)
(where \( \alpha ^{\textrm{v}}\equiv \alpha /\alpha ^{2} \)), one can construct
\( Z \)-monopoles \cite{B78}. The U(1) magnetic charge of these monopoles
are \cite{B78}
\begin{equation}
\label{3.6}
g\equiv \frac{1}{|v|}\int dS_{i}M^{a}B_{i}^{a}=\frac{4\pi }{e}\frac{v\cdot \alpha ^{\textrm{v}}}{|v|}
\end{equation}
where \( v\equiv b\lambda _{\phi } \) and \( B_{i}^{a}\equiv -\epsilon _{ijk}G_{jk}^{a}/2 \)
are the non-Abelian magnetic fields. These monopoles fill supermultiplets of
\( N=2 \) supersymmetry \cite{Adda} and satisfy the mass formula 
\begin{equation}
\label{3.6a}
m_{\textrm{mon}}=|v||g|\, .
\end{equation}
Not all of these monopoles are stable. The stable or fundamental BPS monopoles
are those lowest magnetic charge associated to the roots \( \alpha  \) which
satisfy \( 2\alpha ^{\textrm{v}}\cdot \lambda _{\phi }=\pm 1 \)\cite{EWeinberg}.
From now on, we shall only consider these fundamental monopoles, which are believed
to fill representations of the gauge subgroup \( K \)\cite{HolloDoFraMACK}.
Their magnetic charge (\ref{3.6}) can be written as
\begin{equation}
\label{3.7}
g=\frac{2\pi k}{q_{\phi }}
\end{equation}

Differently from the case in which the gauge group is broken to \( U(1)^{\textrm{rank}(G)} \),
it is very important to note that when \( G \) is broken to \( U(1)\times K/Z_{l} \),
the existence of monopole solutions can explain the 1/3 factor in the electric
charge quantization condition for the colored particles like quarks and gluons
when \( K=SU(3) \) \cite{CorriganOlive}.

It is interesting to note that for the particular case where the gauge group
is \( G=SU(2) \) and scalar \( \phi  \) belongs to symmetric part of the direct
product of the fundamental with itself, which corresponds to the adjoint representation,
the supersymmetry of the theory is enhanced to \( N=4 \). In this case, the
theory has vanishing beta function. 

There are other examples of vanishing \( \beta  \) functions when \( m=0 \).
The \( \beta  \) function of \( N=2 \) super Yang-Mills with a hypermultiplet
is given by
\[
\beta (e)=\frac{-e^{3}}{\left( 4\pi \right) ^{2}}\left[ h^{\textrm{v}}-x_{\phi }\right] \]
where \( h^{\textrm{v}} \) is the dual Coxeter number of \( G \) and \( x_{\phi } \)
is the Dynkin index of \( \phi  \)'s representation. If \( \phi  \) belongs
to the direct product of \( 2 \) fundamental representations, \( R^{\otimes }_{2\lambda _{\phi }} \),
\[
x_{\phi }=2d_{\lambda _{\phi }}x_{\lambda _{\phi }}\, ,\]
where \( x_{\lambda _{\phi }} \) and \( d_{\lambda _{\phi }} \) are, respectively,
the Dynkin index and the dimension of the representation associated to the fundamental
weight \( \lambda _{\phi } \). 

Therefore for \( SU(N) \) (which has \( h^{\textrm{v}}=N \)), if \( \phi  \)
is in the tensor product of the fundamental representation of dimension \( d_{\lambda _{N-1}}=N \)
with itself (which has Dynkin index \( x_{\lambda _{N-1}}=1/2 \)), \( x_{\phi }=N \)
and the \( \beta  \) function vanishes. Therefore the theory is superconformal
(if we take \( \mu =0 \)). In this phase, \( SU(N) \) is broken to \( U(N-1)\sim \left( SU(N-1)\otimes U(1)\right) /Z_{N-1} \).

\section{The \protect\( m>0\protect \) or superconducting phase}

In the ``\( m>0 \)'' phase, the U(1) factor is broken and, like in the Abelian-Higgs
theory, the corresponding force lines cannot spread over space. Since \( G \)
is broken in such a way that \( \Pi _{1}(G/G_{\phi })=Z_{k} \), these force
lines may form topological \( Z_{k}- \)strings. We shall show the existence
of BPS \( Z_{k} \)-strings in this phase and obtain their string tensions.
We shall also show that, as in the Abelian Higgs theory, the U(1) magnetic flux
\( \Phi _{\textrm{mon}} \) of the above monopoles is proportional to the BPS
\( Z_{k} \)-string magnetic flux \( \Phi _{\textrm{st}} \), and therefore
these \( U(1) \) flux lines coming out of the monopole can be squeezed into
\( Z_{k} \)-strings, which can give rise to a confining potential.

\subsection{The BPS \protect\( Z_{k}\protect \)-string solutions}

At \( \rho \rightarrow \infty  \), the string must tend to vacuum solutions
in any angular direction \( \varphi  \). Let us denote \( \phi (\varphi )=\phi (\varphi ,\rho \rightarrow \infty ), \)
\( S(\varphi )=S(\varphi ,\rho \rightarrow \infty ) \), etc. Then, the vacuum
equations (\ref{7.3}) imply that this asymptotic field configuration must be
related by gauge transformations from a vacuum configuration, which we shall
consider (\ref{2.5b}), i.e.
\begin{eqnarray*}
W_{I}(\varphi ) & = & \frac{-1}{ie}\left( \partial _{I}g(\varphi )\right) g(\varphi )^{-1}\, \, ,\, \, \, \, \, \, \, I=1,2\, ,\\
\phi (\varphi ) & = & g(\varphi )\phi ^{\textrm{vac}}\, ,\\
S(\varphi ) & = & g(\varphi )S^{\textrm{vac}}g(\varphi )^{-1}\, ,
\end{eqnarray*}
for some \( g(\varphi )\in G \). Then, in order for the field configurations
to be single-valued, \( g(2\pi )g(0)\in G_{\phi } \). Without lost of generality
we shall consider \( g(0)=1 \). Since \( G \) is simply connected (which can
always be done by going to the universal covering group), a necessary condition
for the existence of strings is that \( g(2\pi ) \) belongs to a non-connected
component of \( G_{\phi } \). Let \( g(\varphi )=\exp i\varphi L \). Then,
at \( \rho \rightarrow \infty  \) 
\begin{eqnarray}
\phi (\varphi ) & = & ae^{i\varphi L}|k\lambda _{\phi }>\, ,\nonumber \\
mS(\varphi ) & = & ka^{2}e^{i\varphi L}\lambda _{\phi }\cdot He^{-i\varphi L}\, ,\label{3.4} \\
W_{I}(\varphi ) & = & \frac{\epsilon _{IJ}x^{J}}{e\rho ^{2}}L\, ,\, \, \, \, \, \, \, I,J=1,2\, .\nonumber 
\end{eqnarray}
Some possible choices for the Lie algebra element \( L \) are
\begin{equation}
\label{3.4a}
L_{n}=\frac{n}{k}\frac{\lambda _{\phi }\cdot H}{\lambda _{\phi }^{2}}\, ,
\end{equation}
with \( n \) being a non-vanishing integer defined modulo \( k \). With these
choices it is possible to show\cite{KB} that \( g(2\pi )\in G_{\phi } \). 

We can construct the string ansatz by multiplying by arbitrary functions of
\( \rho  \), the asymptotic configuration (\ref{3.4}) with \( L_{n} \) given
by (\ref{3.4a}), 
\begin{eqnarray}
\phi (\varphi ,\rho ) & = & f(\rho )e^{in\varphi }a|k\lambda _{\phi }>\, ,\nonumber \\
mS(\varphi ,\rho ) & = & h(\rho )ka^{2}\lambda _{\phi }\cdot H\, ,\label{4.4} \\
W_{I}(\varphi ,\rho ) & = & g(\rho )L_{n}\frac{\epsilon _{IJ}x^{J}}{e\rho ^{2}}\, \, \, \, \, \, \, \rightarrow \, \, \, \, \, \, \, B_{3}(\varphi ,\rho )=\frac{L_{n}}{e\rho }g'(\rho )\, ,\nonumber \\
W_{0}(\varphi ,\rho ) & = & W_{3}(\varphi ,\rho )=0\, ,\nonumber 
\end{eqnarray}
with the boundary conditions
\[
f(\infty )=g(\infty )=h(\infty )=1\, ,\]
in order to recover the asymptotic configuration at \( \rho \rightarrow \infty  \)
and 
\[
f(0)=g(0)=0\]
in order to eliminate singularities at \( \rho =0 \).

Putting this ansatz in the BPS conditions it results that:
\begin{eqnarray*}
h(\rho ) & = & \textrm{const }=1\\
f'(\rho ) & = & \pm \frac{n}{\rho }\left[ 1-g(\rho )\right] f(\rho )\\
g'(\rho ) & = & \mp \frac{q_{\phi }^{2}a^{2}\rho }{2n}\left[ |f(\rho )|^{2}-1\right] 
\end{eqnarray*}
which are exactly the same differential equations with same boundary conditions
which appear in the \( U(1) \) case (\ref{0.4}) (with \( q_{\phi } \) given
by (\ref{4.1})), whose existence of solution has been proven by Taubes. 

As we mentioned before, the BPS conditions are compatible with the equations
of motion when \( m \) vanishes. However, if we do this, \( a=0 \) and there
is no symmetry breaking, which is necessary in order for string solutions to
exist. This result is very similar to what happens for the BPS monopole (see
for instance \cite{GOrev}). In that case, one of the BPS conditions is \( V(\phi )=\lambda (\phi ^{2}-a^{2})^{2}/4=0 \),
which implies the vanishing of the coupling \( \lambda  \). However, that condition
must be understood in the Prasad-Sommerfield limiting case \( \lambda \rightarrow 0 \)
\cite{PrasadSommerfield} in order to retain the boundary condition \( |\phi |\rightarrow a \)
as \( r\rightarrow \infty  \), and to have symmetry breaking. In our case,
we have the same situation with a small difference: if one considers \( m\rightarrow 0 \),
then \( a\rightarrow 0 \). We can avoid this problem by allowing \( \mu \rightarrow \infty  \)
such that \( m\mu  \), or equivalently \( a \), remains constant, implying
that the field \( \phi  \) becomes infinitely heavy.

\subsection{The \protect\( Z_{k}\protect \)-string magnetic flux and the string tension}

Since the scalar \( M^{a} \) defines the \( U(1) \) direction inside \( G \),
we define a gauge invariant \( U(1) \) string magnetic flux
\begin{equation}
\label{4.3}
\Phi _{\textrm{st}}\equiv \frac{1}{|v|}\int d^{2}x\, M^{a}B^{a}_{3}
\end{equation}
 which is similar to the \( U(1) \) monopole magnetic flux (\ref{3.6}), but
with surface integral taken over the plane perpendicular to the string. Using
the BPS string ansatz we obtain that 
\begin{equation}
\label{4.4c}
\Phi _{\textrm{st}}=\oint dl_{I}A_{I}=\frac{2\pi n}{q_{\phi }}\, \, \, \, ,\, \, \, \, \, \, \, \, \, \, n\in Z_{k\, },
\end{equation}
where \( A_{I}\equiv W_{I}^{a}M^{a}/|v|\, ,\, \, \, I=1,2 \) and \( q_{\phi } \)
given by (\ref{4.1}). This flux quantization condition is also very similar
to the Abelian result\footnote{%
It is interesting to note that (\ref{4.4c}) corresponds to the Abelian flux
(\ref{fluxo}) divided by \( |\lambda _{\phi }| \) when \( k=2 \) .
} (\ref{fluxo}) and generalizes, for example, the string magnetic flux for \( SU(2) \)\cite{Schaposnik}
and for \( SO(10) \)\cite{Everett}(up to a \( \sqrt{2} \) factor). In \cite{konishi},
it is also calculated fluxes for the \( SU(n) \) theory, but with the gauge
group completely broken to its center and a different definition of string flux
which is not gauge invariant. Note that we can rewrite the above result as
\[
\Phi _{\textrm{st}}q_{\phi }=2\pi n\, \, ,\, \, \, \, \, \, n\in Z_{k}\, .\]

We can conclude that for the fundamental (anti)monopoles, the \( U(1) \) magnetic
flux \( \Phi _{\textrm{mon}}=g \), which is given in (\ref{3.7}), is consistent
with \( \Phi _{\textrm{st}} \), if \( n=k \). This can be interpreted that
for one fundamental monopole we could attach \( k \) \( Z_{k} \)-strings with
\( n=1 \). That is consistent with the fact that \( k \) \( Z_{k}- \)strings
with \( n=1 \) have trivial first homotopy, as do the monopoles. 

With the above definition of string flux, the string tension bound (\ref{7.5})
can be written as
\[
T\geq \frac{1}{2}q_{\phi }|a|^{2}|\Phi _{\textrm{st}}|=\pi |a|^{2}|n|\, ,\, \, \, n\in Z_{k}\, ,\]
which generalizes the \( U(1) \) results (\ref{2.3a}) and (\ref{1.4}). The
bound hold for the BPS string. Since the tension is constant, it may cause a
confining potential between monopoles increasing linearly with their distance,
which may produce quark confinement in a dual theory.

\subsection{The monopole confinement}

In the \( m<0 \) phase, by topological arguments \cite{B81}\cite{PV} one
would expect that the monopoles produced in the \( m=0 \) phase develop a flux
line or string and get confined. We can see this more concretely in the following
way: as usual, in order to obtain the asymptotic scalar configuration of a (spherically
symmetric) monopole, starting from the vacuum configuration (\ref{2.5b}) one
performs a spherically symmetric gauge transformation. Then for the case of
a fundamental monopole and considering the \( k=2 \) we can show\cite{Kneipp2002}
that asymptotically 

\[
\phi (\theta ,\varphi )=a\left\{ \cos ^{2}\frac{\theta }{2}|2\lambda _{\phi }>-\frac{\sqrt{2}}{2}\sin \theta e^{-i\varphi }|2\lambda _{\phi }-\alpha >+\sin ^{2}\frac{\theta }{2}e^{-2i\varphi }|2\lambda _{\phi }-2\alpha >\right\} \, .\]
Therefore at \( \theta =\pi  \),
\[
\phi (\pi ,\varphi )=ae^{-2i\varphi }|2\lambda _{\phi }-2\alpha >\]
 which is singular. This generalizes Nambu's result \cite{Nambu} for the \( SU(2)\times U(1) \)
case. In order to cancel the singularity we should attach a string in the \( z<0 \)
axis with a zero in the core, as in our string ansatz (\ref{4.4}). One could
construct an ansatz for \( \phi (r,\theta ,\phi ) \) by multiplying the above
asymptotic configuration by a function \( F(r,\theta ) \) such that \( F(r,\pi )=0 \).

The string tension for \( k \) strings with \( n=1 \) must satisfy
\[
T\geq k\pi |a|^{2}\, .\]
 Then, the threshold length \( d^{\textrm{th}} \) for the string to break producing
a new monopole-antimonopole pair, with masses (\ref{3.6a}), is derived form
the relation
\[
\frac{4\pi }{e}\frac{k|a|^{2}|\lambda _{\phi }|}{m}=E^{\textrm{th}}=Td^{\textrm{th}}\geq k\pi |a|^{2}d^{\textrm{th}}\, ,\]
which results in 
\[
d^{\textrm{th}}\leq \frac{4|\lambda _{\phi }|}{me}\, .\]
 The monopole-antimonopole pair tends to deconfine when \( m\, \rightarrow \, 0_{+} \),
as one would expect, when \( T\rightarrow 0 \) and \( d^{\textrm{th}}\rightarrow \infty  \).

\section{Summary and conclusions}

In this talk, we have presented some generalizations of the ideas of 't Hooft
and Mandelstam to non-Abelian theories. Like in the Abelian-Higgs theory, we
have seen that our non-Abelian theory presents some different phases, including
a superconducting phase, where we have proven the existence of BPS \( Z_{k} \)-string
solutions and have calculated exactly their string tension. We also showed that
the fluxes of the magnetic monopoles and strings are proportional and therefore
the monopoles can get confined. But differently from the Abelian case, or also
from the theories in which are used Abelian projection, our monopoles are not
Dirac monopoles, and so we can associate naturally mass to them. Due to that,
we could calculate the threshold length for the string to break in a new monopole-antimonopole
pair. Also in our theory the unbroken gauge group is non-Abelian and our monopoles
should fill some representation of this unbroken gauge group. In many cases,
the unbroken group can contain \( SU(3)\otimes U(1)_{\textrm{em}} \) as subgroup.
Moreover we have seen that we could consider the scalar \( \phi  \) as a diquark
condensate and unlike the Abelian theory, in our theory the bare mass \( \mu  \)
of \( \phi  \) is not required to satisfy \( \mu ^{2}<0 \) in order to have
spontaneous symmetry breaking. Therefore in the dual formulation, where one
could interpret \( \phi  \) as being the monopole condensate, we don't need
to have a monopole mass satisfying the problematic condition \( M_{\textrm{mon}}^{2}<0 \)
mentioned by 't Hooft.

It is expected that a confining theory obtained by a deformation of superconformal
gauge theory in 4 dimensions should satisfy a gauge/string correspondence \cite{Witten},
which would be a kind of deformation of the CFT/AdS correspondence \cite{Maldacena}.
In the gauge/string correspondences it is usually considered confining gauge
theories with \( SU(N) \) or \( U(N) \) completely broken to its center. We
have seen that some of our confining theories are obtaining by adding a deformation
to superconformal theories and which breaks \( U(N-1) \) to \( SU(N-1)\otimes Z_{2} \)
(up to a discrete factor). It would be interesting to know if these theories
also satisfy a gauge/string correspondence.

\acknowledgments

I would like to thank Patrick Brockill for collaboration and José Helayel for
reading the manuscript. I also would like thank the organizers of the workshop
for the kind hospitality.

\end{document}